\documentclass[aps,prd,preprintnumbers,superscriptaddress,nofootinbib]{revtex4}
\usepackage[dvips]{graphicx}
\usepackage{bm,latexsym,amsmath,amssymb,amsfonts,mathrsfs}
\usepackage{color}
\input{colordvi.tex}
\newcommand*{\D}{{\rm d}}
\newcommand*{\mpl}{M_{\rm Pl}}
\begin{document}

\title{Black hole perturbation in the most general scalar-tensor theory with
second-order field equations I:~The odd-parity sector}

\author{Tsutomu~Kobayashi}
\email[Email: ]{tsutomu"at"tap.scphys.kyoto-u.ac.jp}
\affiliation{Hakubi Center, Kyoto University, Kyoto 606-8302, Japan
}
\affiliation{Department of Physics, Kyoto University, Kyoto 606-8502, Japan
}

\author{Hayato~Motohashi}
\email[Email: ]{motohashi"at"resceu.s.u-tokyo.ac.jp}
\affiliation{Department of Physics, Graduate School of Science,
The University of Tokyo, Tokyo 113-0033, Japan
}
\affiliation{Research Center for the Early Universe (RESCEU),
Graduate School of Science,
The University of Tokyo, Tokyo 113-0033, Japan
}

\author{Teruaki~Suyama}
\email[Email: ]{suyama"at"resceu.s.u-tokyo.ac.jp}
\affiliation{Research Center for the Early Universe (RESCEU),
Graduate School of Science,
The University of Tokyo, Tokyo 113-0033, Japan
}

\begin{abstract}
We perform a fully relativistic analysis of
odd-type linear perturbations around a static and spherically symmetric
solution in the most general scalar-tensor theory with second-order
field equations in four-dimensional spacetime.
It is shown that, as in the case of general relativity, 
the quadratic action for the perturbations reduces to
the one having only a single dynamical variable, from which
concise formulas for no-ghost and no-gradient instability conditions are derived.
Our result is applicable to all the theories of gravity
with an extra scalar degree of freedom.
We demonstrate how the generic formulas can be applied to
some particular examples such as the Brans-Dicke theory, $f(R)$ models,
and Galileon gravity.
\end{abstract}

\preprint{RESCEU-3/12, KUNS-2385}
\maketitle

\section{Introduction}

General relativity is definitely the most successful theory
of gravity, passing all the experimental and observational tests performed so far.
Nevertheless, it is still important to ask, ``To what extent was Einstein correct?''
and motivated by this question
a number of alternative theories have been proposed.
In particular,
since the discovery of the cosmic acceleration~\cite{sne},
the mystery of dark energy has led us to explore more seriously the possibility of
modifying gravity on cosmological scales.

In order for a given modified gravity model to be viable,
it must pass stringent tests in the solar system and on the Earth.
The standard cosmological evolution after
nucleosynthesis must be reproduced as well in modified gravity.
In addition to those requirements,
one would demand that, as a necessary condition to ensure the stability of our world, the static and spherically symmetric metric is stable.
The purpose of this paper is
to provide the stability criteria against
linear perturbations about a static and spherically symmetric vacuum solution
in modified gravity,
focusing, as a first step, on odd-parity modes.
This is done by computing the quadratic action for
the odd-parity perturbations.
We try to make our results as general as possible in such a way that
the formulas we present are applicable to
all the modified gravity theories having
a single scalar degree of freedom $\phi$ on top of the metric $g_{\mu\nu}$.
To this end, we employ the most general scalar-tensor theory
with second-order field equations developed originally by Horndeski~\cite{Horndeski}.
Horndeski's theory contains novel models that have never been studied,
as well as the Brans-Dicke theory, $f(R)$ models~\cite{frreview}, and
Galileon gravity~\cite{Galileon, CovGali} as special cases.
Therefore, our results are a useful and powerful discriminant to judge
whether or not a given modified gravity model is viable
according to the stability of a black hole.

We closely follow the approach of Refs.~\cite{Motohashi:2011pw,Motohashi:2011ds}
for black hole perturbations,
in which two of the authors studied black hole perturbations in
a gravitational theory whose
Lagrangian is a general function $f(R,C)$
of the Ricci scalar $R$ and
the Chern-Simons combination $C$.

The paper is organized as follows.
In the next section, we introduce the
Lagrangian for the most general second-order scalar-tensor theory.
After providing the background
field equations in Sec.~III,
we derive in Sec.~IV the quadratic action for odd-parity perturbations
and a master equation for a single dynamical variable.
From the quadratic action we give
concise formulas for no-ghost and no-gradient instability conditions.
Our generic formulas are then applied
to some representative models of modified gravity in Sec.~V.
Finally, we draw our conclusion in Sec.~VI.

Throughout the paper, we work in the natural unit $c=1$, and our sign conventions are: $\eta_{\mu\nu}={\rm diag}(-1,+1,+1,+1)$, $R^\alpha_{~\beta \gamma \delta}=\partial_\gamma \Gamma^\alpha_{~\beta \delta}-\partial_\delta \Gamma^\alpha_{~\beta \gamma}+\Gamma^\alpha_{~\lambda \gamma} \Gamma^\lambda_{~\beta \delta}-\Gamma^\alpha_{~\lambda \delta} \Gamma^\lambda_{~\beta \gamma}$ and $R_{\mu \nu}=R^\alpha_{~\mu \alpha \nu}$.

\section{The most general second-order scalar-tensor theory}

We consider the most general scalar-tensor theory
having second-order field equations both for the metric and the scalar field
in four spacetime dimensions.
The theory was originally determined by Horndeski~\cite{Horndeski},
and later it was rederived in Ref.~\cite{GG} as the most general
extension of the Galileon field~\cite{Galileon}. The equivalence of
Horndeski's theory~\cite{Horndeski} and the generalized Galileon~\cite{GG}
was shown explicitly in Ref.~\cite{G2}.
The generalized Galileon is described by the following four
Lagrangians:
\begin{eqnarray}
{\cal L}_2&=&K(\phi, X),
\\
{\cal L}_3&=&-G_3(\phi, X)\Box\phi,
\\
{\cal L}_4&=&G_4(\phi, X)R+G_{4X}\left[(\Box\phi)^2-(\nabla_\mu\nabla_\nu\phi)^2\right],
\\
{\cal L}_5&=&G_5(\phi, X)G_{\mu\nu}\nabla^\mu\nabla^\nu\phi
-\frac{1}{6}G_{5X}\bigl[(\Box\phi)^3
-3\Box\phi(\nabla_\mu\nabla_\nu\phi)^2+
2(\nabla_\mu\nabla_\nu\phi)^3\bigr],
\end{eqnarray}
where $K$ and $G_i$ are arbitrary functions of $\phi$ and $X:=-(\partial\phi)^2/2$.
Here, we used the notation $G_{iX}$ for $\partial G_i/\partial X$.
We thus want to consider the action
\begin{eqnarray}
S=\sum_{i=2}^5\int\D^4x\sqrt{-g}{\cal L}_i. \label{action1}
\end{eqnarray}

Horndeski's theory includes
all the dark energy and modified gravity models with
a single scalar degree of freedom in addition to the metric tensor.
For example, ${\cal L}_2$ corresponds to the k-essence field.
Since the nonminimal coupling of the form $F(\phi)R$ can be reproduced
by choosing $G_4(\phi, X)=F(\phi)$, the Brans-Dicke theory and $f(R)$ gravity
are also special cases of the above theory.
The covariant Galileon~\cite{CovGali} corresponds to
$K=-c_2X, G_3=-c_3X/M^3, G_4=\mpl^2/2-c_4X^2/M^6$,
and $G_5=3c_5X^2/M^9$, while the nonminimal coupling to
the Gauss-Bonnet term, $\xi(\phi)\left(R^2-4R_{\mu\nu}^2+R_{\mu\nu\rho\sigma}^2\right)$, to
$
K=8\xi^{(4)}X^2\left(3-\ln X\right),
G_3=4\xi^{(3)}X\left(7-3\ln X\right),
G_4=4\xi^{(2)}X\left(2-\ln X\right),
G_5=-4\xi^{(1)} \ln X,
$ where $\xi^{(n)}:=\partial^n\xi/\partial\phi^n$~\cite{G2}.
The coupling of the form $G^{\mu\nu}\nabla_\mu\phi\nabla_\nu\phi$
introduced {\em e.g.}, in Refs.~\cite{gpp, gpp2}
can be obtained from $G_5=-\phi$ after integration by parts.

Recently, Horndeski's theory has been revived and studied extensively
in the context of cosmology~\cite{G2, fab4, Gao, Gao2, Tsujikawa, Rampei}.
We are going to study Horndeski's theory
from a different viewpoint,
applying this powerful framework
to the perturbation analysis in spherically symmetric spacetime.

\section{Background}

We consider a static and spherically symmetric
background, so that the unperturbed metric $\bar g_{\mu\nu}$ may be written as
\begin{eqnarray}
\bar g_{\mu\nu}\D x^\mu\D x^\nu
=-A(r)\D t^2+\frac{\D r^2}{B(r)}+C(r) r^2\left(\D\theta^2+\sin^2\theta\,\D \varphi^2\right),
\end{eqnarray}
and the scalar field is dependent on the radial coordinate only: $\phi=\phi(r)$
(and therefore $X=-B(\phi')^2/2$, where a prime stands for
differentiation with respect to $r$.)
Probably the easiest way to derive the background field equations
is to substitute the above ansatz to the action and to carry out the variation
with respect to $A, B, C$, and $\phi$.
Actually, the metric function $C(r)$ is redundant,
but the introduction of $C(r)$ helps one derive the angular component of the
gravitational field equations.
Of course, we can set $C(r)=1$ without loss of generality after getting
all the field equations.
The background field equations are thus given by
\begin{eqnarray}
{\cal E}_A=0,
\quad
{\cal E}_B=0,
\quad
{\cal E}_C=0,
\quad
{\cal E}_\phi=\frac{1}{r^2}\sqrt{\frac{B}{A}}\frac{\D}{\D r}\left(r^2\sqrt{AB}\,{\cal J}\right)
-{\cal S}=0,
\label{bgeqs}
\end{eqnarray}
where explicit form of ${\cal E}_A, {\cal E}_B, \cdots$ etc.\ is
provided in the appendix.
The functions $A(r),~B(r)$ and $\phi(r)$ may be then determined by
solving the field equations supplemented with appropriate boundary conditions.
However, since we are interested in the generic structure of
the perturbation equations and stability conditions rather than
the properties of possible static and spherically symmetric solutions,
we leave $A, B$, and $\phi$ unspecified and
proceed to the perturbation analysis.
Note in particular that
the background may not be an asymptotically flat solution.

\section{Perturbations}

Given the background equations of motion,
we can now derive the quadratic action for perturbations.
As is well known as the Regge-Wheeler formalism~\cite{Regge:1957td,Zerilli:1970se},
metric perturbations as well as the scalar field perturbation on a
static and spherically symmetric background can be 
decomposed into odd- and even-type perturbations
according to their transformation properties under
a two-dimensional rotation.
Such a decomposition is useful because
the linear perturbation equations for odd and even modes
are completely decoupled from each other,
allowing one to study each mode separately.
In this paper, we focus on the odd-type perturbations, in which case the scalar field does not acquire perturbation.
We will obtain very concise formulas for the odd modes
even in the most general case, while
the computation of the even modes would be quite messy.
A thorough analysis of the even-type perturbations will therefore be
presented elsewhere~\cite{even}.

Each perturbation variable can be decomposed further into the sum of
the spherical harmonics $Y_{\ell m}(\theta,\varphi)$.
Then, at linear order in perturbations, or, equivalently, in the quadratic
action, the perturbation variables having different $\ell$ and $m$
do not mix, which simplifies our perturbation analysis drastically.
Here we only summarize the definition and notation of the perturbation variables,
referring the readers to Refs.~\cite{Motohashi:2011pw,Motohashi:2011ds}
for a more detailed explanation of the Regge-Wheeler formalism.

The odd-type metric perturbations $h_{\mu\nu}=g_{\mu\nu}-\bar g_{\mu\nu}$ can be written as
\begin{eqnarray}
 &  & h_{tt}=0,~~~h_{tr}=0,~~~h_{rr}=0,\\
 &  & h_{ta}=\sum_{\ell, m}h_{0,\ell m}(t,r)E_{ab}\partial^{b}Y_{\ell m}(\theta,\varphi),\\
 &  & h_{ra}=\sum_{\ell, m}h_{1,\ell m}(t,r)E_{ab}\partial^{b}Y_{\ell m}(\theta,\varphi),\\
 &  & h_{ab}=\frac{1}{2}\sum_{\ell, m}h_{2,\ell m}(t,r)\left[E_{a}^{~c}\nabla_{c}\nabla_{b}Y_{\ell m}(\theta,\varphi)+E_{b}^{~c}\nabla_{c}\nabla_{a}Y_{\ell m}(\theta,\varphi)\right], \label{odd-ab}
\end{eqnarray}
where $E_{ab}:=\sqrt{\det\gamma}~\epsilon_{ab}$
with $\gamma_{ab}$ being the two-dimensional metric on the sphere and $\epsilon_{ab}$ being the totally antisymmetric symbol with $\epsilon_{\theta \varphi}=1$.

Because of general covariance, not all the metric perturbations are physical 
in the sense that some of them can be set to vanish by using the gauge 
transformation $x^{\mu}\to x^{\mu}+\xi^{\mu}$, where $\xi^{\mu}$ is an
infinitesimal function.
For the odd-type perturbations, one can completely fix the gauge by imposing
the condition $h_2=0$ for $\ell \ge 2$.
This is called the Regge-Wheeler gauge, which we will use
in the following calculations.
For the dipole perturbation $\ell=1$, $h_{ab}$ identically vanishes and
we need to impose another gauge condition.
We will therefore treat this case separately following the study for the general case.

\subsection{Second-order action}

Expanding the action (\ref{action1}) to second order in the perturbations,
we obtain the following action:
\begin{eqnarray}
S^{(2)}=\int \D t\,\D r\,{\cal L}^{(2)},
\end{eqnarray}
where ${\cal L}^{(2)}$ is of the form
\begin{equation}
\frac{2\ell+1}{2\pi} {\cal L}^{(2)}=a_1 h_0^2+a_2 h_1^2+a_3 \left( {\dot h_1}^2-2 {\dot h_1} h_0'+h_0'^2+\frac{4}{r} {\dot h_1} h_0 \right), \label{lag-odd1}
\end{equation}
where a dot denotes differentiation with respect to $t$.
Since modes with different $(\ell,\,m)$ do not mix,
we have picked up the perturbations belonging to a given $(\ell,\,m)$.
Thanks to the spherical symmetry of background spacetime,
the action for the $m \neq 0$ modes takes exactly the same form as that for $m=0$,
and hence we may set $m=0$ without loss of generality.
Here and hereafter we abbreviate the subscripts $\ell$ and $m$.
The coefficients $a_1, a_2$, and $a_3$ in the above Lagrangian are given by
\begin{eqnarray}
a_1&=&\frac{\ell (\ell+1)}{r^2}
\left[\frac{d}{d r}\left(
r\sqrt{\frac{B}{A}}{\cal H}\right)
+\frac{\ell^2+\ell-2}{2\sqrt{AB}}{\cal F} 
+\frac{r^2}{\sqrt{AB}} {\cal E}_A \right],
\\
a_2&=&- \ell (\ell+1) \sqrt{AB} \left[ \frac{(\ell-1) (\ell+2)}{2r^2}{\cal G}+{\cal E}_B \right],
\\
a_3&=&
\frac{\ell (\ell+1)}{2} \sqrt{\frac{B}{A}} {\cal H},
\end{eqnarray}
where 
\begin{eqnarray}
{\cal F}&:=&2\left(
G_4+\frac{1}{2}B\phi'X'G_{5X}-XG_{5\phi}
\right), \label{def-f}
\\
{\cal G}&:=&2\left[ G_4-2XG_{4X}
+X\left(\frac{A'}{2A}B\phi'G_{5X}+G_{5\phi}\right)
\right], \label{def-g}
\\
{\cal H}&:=&2\left[G_4-2X G_{4X}
+X\left(\frac{B \phi'}{r} G_{5X}+ G_{5\phi}\right)\right]=-\frac{2A}{B} \frac{\partial {\cal E}_C}{\partial A''}. \label{def-h}
\end{eqnarray}
Note that the background field equations imply ${\cal E}_A=0={\cal E}_B$,
allowing us to make these coefficients a little simpler.
In the above equations, we have left ${\cal E}_A$ and ${\cal E}_B$ as they are
in order to clarify the places where the background equations of motion
are used.

Notice that only $G_4$ and $G_5$ appear in the coefficients.
This is also the case in tensor cosmological perturbations
(corresponding to gravitational waves) at first and second order~\cite{G2, Gao2}.
Given that the odd modes represent tensor degrees of freedom,
it is reasonable to have such a similarity between perturbations on the different backgrounds.
One may also notice that $G_5$ appears only in the form of derivatives,
which is consistent with the fact
that ${\cal L}_5$ reduces to a total derivative if $G_5={\rm const}$.
Let us also remark on the cases when all of or some of ${\cal F}, {\cal G}$, and ${\cal H}$ happen to vanish. For the case of ${\cal H}=0$, ${\cal F}\not=0$, ${\cal G}\not=0$, the odd-type perturbations vanish: $h_0=0=h_1$. For the other cases, we cannot determine both $h_0$ and $h_1$ completely.

Since ${\dot h_0}$ is absent in the quadratic Lagrangian~(\ref{lag-odd1}),
$h_0$ is an auxiliary field.
However, the presence of the term $h_0'^2$ hinders removing $h_0$ from the
Lagrangian in a straightforward manner.
To eliminate the nondynamical degree of freedom, we invoke the following trick~\cite{DeFelice:2011ka}.
Let us rewrite Eq.~(\ref{lag-odd1}) as
\begin{equation}
\frac{2\ell+1}{2\pi} {\cal L}^{(2)}=
\left( a_1-\frac{2(ra_3)'}{r^2} \right) h_0^2
+a_2 h_1^2+a_3
\left[ -q^2+2q \left( {\dot h_1}- h_0'+\frac{2}{r} h_0 \right) \right]. \label{lag-odd2}
\end{equation}
It is easy to verify that the new Lagrangian~(\ref{lag-odd2})
reduces to the original one~(\ref{lag-odd1}) after eliminating 
the auxiliary field $q$.
The new Lagrangian enables
us to put all the derivatives on $q$ by integration by parts, transforming both $h_0$
and $h_1$ to removable auxiliary fields.
Variation with respect to $h_0$ and $h_1$ now yields the following algebraic equations
respectively:
\begin{equation}
h_0=-\frac{r \left[r a_3' q+a_3 \left(r q'+2 q\right)\right]}{r^2 a_1-2 \left(r a_3'+a_3\right)},~~~h_1=\frac{a_3}{a_2} {\dot q}.\label{hhq}
\end{equation}
Plugging this result back into Eq.~(\ref{lag-odd2}), we obtain
the quadratic Lagrangian solely in terms of the master variable $q$.
Once solving for $q$, one can determine the metric perturbations $h_0$ and $h_1$
through the relation~(\ref{hhq}).

To arrive at our final result, it is convenient to rescale the variable $q$ as
\begin{equation}
Q={\left( \frac{r^2 B^{3/2} {\cal H}^2}{\sqrt{A} {\cal F}} \right)}^{1/2} q.
\label{Qq}
\end{equation}
With this rescaling,
the coefficient of the gradient term is transformed to $-1$
and the structure of the final Lagrangian will be of the clearest form.
Here, we implicitly assumed that ${\cal F}>0$
for the field redefinition~(\ref{Qq}) to be feasible. 
Otherwise, the system would exhibit gradient instability.
We thus obtain the quadratic Lagrangian for $Q$:
\begin{equation}
\frac{2\ell+1}{2\pi} {\cal L}^{(2)}=\frac{\ell (\ell+1)}{2 (\ell-1) (\ell+2)}
\left[ \frac{{\cal F}}{AB {\cal G}}{\dot Q}^2-Q'^2
-\frac{\ell (\ell+1) {\cal F}}{r^2 B{\cal H}} Q^2 -V(r) Q^2 \right],
\label{Lfin}
\end{equation}
where the effective potential $V(r)$ is given by
\begin{eqnarray}
V(r)&=&-\frac{1}{16} \bigg[ \frac{A'^2}{A^2} + 4 \frac{A'}{A} \left( \frac{{\cal F}'}{{\cal F}}+\frac{{\cal H}'}{{\cal H}}+\frac{3}{r} \right)+3 \frac{B'^2}{B^2}+4 \left( \frac{B'}{B}\frac{{\cal F}'}{{\cal F}}+\frac{B'}{rB}-\frac{B''}{B}+\frac{10 {\cal F}}{Br^2 {\cal H}} \right)  \nonumber \\
&&-4 \left( 3 \frac{{\cal F}'^2}{{\cal F}^2} -\frac{2 {\cal F}''}{{\cal F}} +\frac{4 {\cal F}'}{r {\cal F}} +\frac{2 {\cal H}'}{r {\cal H}}+\frac{10}{r^2} \right) \bigg].
\end{eqnarray}

Noting that we are imposing ${\cal F} >0$
in order to avoid gradient instability,
we find a very
concise expression for the no-ghost condition:
\begin{equation}
{\cal G}>0.
\end{equation}
The equation of motion derived from
the Lagrangian~(\ref{Lfin}) is
\begin{equation}
\frac{{\cal F}}{AB {\cal G}}{\ddot Q}-Q''+\frac{\ell (\ell+1) {\cal F}}{r^2 B{\cal H}} Q +V(r) Q=0.\label{waveeq}
\end{equation}
Since $\ell (\ell+1)$ corresponds to the two-dimensional
Laplacian in the real space, the third term in Eq.~(\ref{waveeq}) 
represents wave propagation along the angular direction.
Therefore,
the squared propagation speeds along the radial direction, $c_r^2$, and the
angular direction, $c_\theta^2$, are given respectively by
\begin{equation}
c_r^2=\frac{{\cal G}}{{\cal F}},\quad
c_\theta^2=\frac{{\cal G}}{{\cal H}},
\end{equation}
which in general differ from unity.
The positivity of the first one is equivalent to the stability conditions
mentioned above.
To ensure the stability in the angular direction we further impose
\begin{equation}
{\cal H}>0.
\end{equation}
Interestingly,
the speed of gravitational waves depends on the propagation direction
in general second-order scalar-tensor theories.
To summarize, ${\cal F}>0,\,{\cal G}>0$, and ${\cal H}>0$ are necessary, but not sufficient,
conditions for the stability of a static and spherically symmetric solution.

\subsection{Dipole perturbation: $\ell=1$}

As we mentioned earlier, the discussion so far cannot be applied to the $\ell=1$ mode.
This may be also understood from the final Lagrangian~(\ref{Lfin}), which is 
ill-defined for $\ell=1$ due to the overall factor $(\ell-1)^{-1}$.
For $\ell=1$ case one can eliminate $h_1$ by using the residual gauge degree of freedom.
Furthermore, we note that $h_0$ contains a gauge mode, $h_0=C(t)r^2$, where $C(t)$
is an arbitrary function of time. The equation of motion for $h_0$ is given by
\begin{eqnarray}
&& \dot h_0'-\frac{2}{r}\dot h_0=0, \\
&& a_3h_0''+a_3'h_0'-\frac{2(ra_3)'}{r^2}h_0=0, 
\end{eqnarray}
with $a_3={\cal H}\sqrt{B/A}$. The solution is found to be
\begin{equation}
h_0=\frac{3Jr^2}{4\pi}\int^r\frac{\D \tilde r}{\tilde r^4{\cal H}}\sqrt{\frac{A}{B}}+C(t)r^2,
\end{equation}
where $J$ and $C(t)$ are integration constants.
The second term is nothing but the gauge mode and hence can be eliminated
by using the residual gauge degree of freedom. 
The first term cannot be removed by any gauge transformations
and therefore is a physical perturbation.
More specifically,
it corresponds to a metric around a
slowly rotating black hole
and $J$ is related to its angular momentum.
This point is clearly understood by considering the case in general relativity:
in that case the solution reduces to $h_0=-J/(4\pi M_P^2r)$,
which coincides with the Kerr metric expanded to first order in the angular momentum $J$.

\section{Application to particular models}

The results obtained in the previous section are so general 
that
they can be applied to any scalar-tensor theory having second-order field equations.
To demonstrate how useful our formalism is, in this section
we will apply our stability conditions for odd-type perturbations to some particular examples that have been frequently
studied in the literature as potential dark energy and modified gravity models.

\subsection{Nonminimally coupled models}

We begin with a familiar example of nonminimal coupling of the form $F(\phi)R$.
Since $K$ and $G_3$ are uncorrelated with
stability against odd-parity perturbations,
we may leave the two functions arbitrary.
We are thus led to consider the following subclass of
Horndeski's theory:
\begin{equation}
K=K(X,\phi),~G_3=G_3(X,\phi),~G_4=F(\phi),~G_5=0.
\end{equation}
This reduces to general relativity
with a minimally coupled canonical scalar field for the
choice $K=X-V(\phi),\,G_3=0,$ and $F=\mpl^2/2=\,$const, and
to the original Brans-Dicke theory for the choice $K=\omega X/\phi,\, G_3=0,$
and $F=\phi$ with $\omega$ being a constant parameter.
$f(R)$ gravity can be mapped to a scalar-tensor theory
and is reproduced
with the choice $K=f(\phi)-\phi f'(\phi),\,G_3=0,$ and $F=f'(\phi)$.
Kinetic gravity braiding~\cite{KGB, GINF, Yamamoto} corresponds to $F(\phi)=\mpl^2/2$
with arbitrary $K$ and $G_3$.
For this class of theories, we find
\begin{equation}
{\cal F}={\cal G}={\cal H}=2 F(\phi).
\end{equation}
Therefore, no-ghost and no-gradient instability conditions are simply translated to
\begin{equation}
F(\phi)>0.
\end{equation}
The propagation speeds both along radial and angular directions are equal to the speed of light,
$c_r^2=c_\theta^2=1$.

\subsection{Covariant Galileon}

Covariant Galileons correspond to
\begin{equation}
K=-c_2X,~G_3=-c_3 \frac{X}{M^3},~G_4=\frac{\mpl^2}{2}-c_4 \frac{X^2}{M^6},~G_5=3c_5 \frac{X^2}{M^9},
\end{equation}
where $c_2,~c_3,~c_4,$ and $c_5$ are dimensionless parameters
and $M$ is some mass scale.
The stability conditions are given by
\begin{eqnarray}
&&{\cal F}=\mpl^2+\frac{2X^2}{M^6} \left( -c_4 +\frac{3c_5 B' \phi'}{M^3}+\frac{2c_5 B\phi''}{M^3} \right)>0, \\
&&{\cal G}=\mpl^2+\frac{6X^2}{M^6} \left( c_4+\frac{c_5 B A' \phi'}{AM^3} \right)>0, \\
&&{\cal H}=\mpl^2+\frac{6X^2}{M^6} \left( c_4+\frac{2c_5 B \phi'}{M^3r} \right)>0.
\end{eqnarray}
These results are valid for any static and spherically symmetric metric regardless of whether or not one considers a spacetime with a black hole. In particular, in the case of a black hole,
it was recently shown that a static and spherically symmetric black hole
cannot sustain a nontrivial profile of a covariant Galileon~\cite{hage},
implying that $\phi'=0$. Therefore, in the case of covariant Galileon
the stability conditions are in fact trivial: $\mpl^2>0$.

\subsection{Coupling to the Gauss-Bonnet term}

Let us consider a scalar field nonminimally coupled to
the Gauss-Bonnet term, so that the gravitational part of the Lagrangian
we want to study is
\begin{eqnarray}
\frac{\mpl^2}{2}R+\xi(\phi)\left(R^2-4R_{\mu\nu}^2+R_{\mu\nu\rho\sigma}^2\right).
\end{eqnarray}
This Lagrangian is obtained from
\begin{eqnarray}
G_4=\frac{\mpl^2}{2}+4\xi^{(2)}X\left(2-\ln X\right),
\quad
G_5=-4\xi^{(1)}\ln X,
\end{eqnarray}
where $\xi^{(n)}:=\partial^n\xi/\partial\phi^n$,
and we emphasize again that a concrete form of $K$ and $G_3$
is not important for stability against odd-parity perturbations.
In this case the stability conditions read
\begin{eqnarray}
{\cal F}&=&\mpl^2-4\xi^{(1)}\left(B'\phi'+2B\phi''\right)-8\xi^{(2)}B\phi'^2>0,
\\
{\cal G}&=&\mpl^2-4\xi^{(1)}\frac{A'}{A}B\phi'>0,
\\
{\cal H}&=&\mpl^2-8\xi^{(1)}\frac{B\phi'}{r}>0.
\end{eqnarray}

As far as the authors know, the study of the odd-type perturbations around
black hole spacetime 
in this type of theories is provided only in Ref.~\cite{Pani:2009wy},
where a specific form for $\xi (\phi)$ is assumed,
$\xi (\phi)\propto \exp \left(\phi/\mpl \right)$.
This form of $\xi (\phi)$ is, for instance, realized as the low-energy effective theory 
for the heterotic string.
In Ref.~\cite{Pani:2009wy}, the following master equation for the perturbations is derived:
\begin{equation}
u''(r)+\left[ V_{\rm PC}(r)+\omega^2 K(r) \right] u(r)=0,
\end{equation}
where $\omega$ is a frequency of the wave. 
By analyzing this equation, the stability against perturbations is also established.
However, $K(r)$ given in Ref.~\cite{Pani:2009wy} appears to depend on $\omega$,
which prevents us from checking consistency between our result and theirs.
Let us also stress that the no-ghost condition is not provided in Ref.~\cite{Pani:2009wy} and
our stability conditions give further restriction on the scalar-Gauss-Bonnet coupling.

\subsection{Coupling to the Einstein tensor}

A scalar field derivatively coupled to
the Einstein tensor has been studied recently
in literature~\cite{gpp, gpp2, Germani:2011ua},
for which the gravitational part of the Lagrangian is given by
\begin{eqnarray}
\frac{\mpl^2}{2}R+\alpha G^{\mu\nu}\partial_\mu\phi\partial_\nu\phi,
\end{eqnarray}
where $\alpha$ is a parameter having dimension of (mass)$^{-2}$.
This can be reproduced from
\begin{equation}
G_4=\frac{\mpl^2}{2},\quad G_5=-\alpha \phi,
\end{equation}
after integration by parts.
The stability conditions are given by
\begin{eqnarray}
{\cal F}=\mpl^2-\alpha B \phi'^2>0,
\quad
{\cal G}={\cal H}=\mpl^2+\alpha B \phi'^2>0.
\end{eqnarray}
The propagation speeds of gravitational waves along the radial
and angular directions are therefore
\begin{equation}
c_r^2=\frac{\mpl^2+\alpha B \phi'^2}{\mpl^2-\alpha B \phi'^2},
\quad
c_\theta^2=1.
\end{equation}
It can be seen that
gravitational waves propagating along the radial direction
are superluminal if $\alpha >0$ and subluminal if $\alpha <0$.

\subsection{Possible application to higher-dimensional scenarios}

So far we have illustrated how one can apply our generic formulas
to some representative models of modified gravity.
Actually, the stability criteria we have derived
can also be applied to rather different situations than this.
Let us make a brief comment on a possible further application.

Consider Lovelock gravity in $D$ dimensions.
The metric we want to discuss is of the form
\begin{eqnarray}
\D s^2=\bar g_{\mu\nu}(x)\D x^\mu\D x^\nu+e^{2\phi(x)}\hat g_{mn}(y)\D y^m\D y^n,
\end{eqnarray}
where $\bar g_{\mu\nu}(x)$ is the metric of
four-dimensional static and spherically symmetric spacetime,
and $\hat g_{mn}(y)$ of $(D-4)$-dimensional compactified space.
Since the Lovelock action yields second-order
field equations for the metric,
Kaluza-Klein reduction leads to a four-dimensional theory
composed of $\bar g_{\mu\nu}$
and $\phi$ whose equations of motion are of second order.
Therefore, the four-dimensional effective theory must be in
a subclass of Horndeski's theory.
This fact was emphasized in Ref.~\cite{kkred}.
Our generic formula can therefore be used to investigate
stability against odd-parity perturbations
homogeneous along the extra dimensions.
One of concrete examples is
five-dimensional black strings in Gauss-Bonnet gravity;
one can study the stability of the solution obtained in Ref.~\cite{bsgb}
by using the results in this paper.

\section{Conclusion}

In this paper, we have studied odd-parity
perturbations around static and spherically symmetric
background spacetime in the most general scalar-tensor theory with second-order
field equations.
Only tensor-type perturbations are excited in the odd-parity sector.
It was shown, by
explicit reduction of the quadratic action using the constraint equations,
that
there is a single master variable
obeying a second-order differential equation,
from which the dynamical
behavior of all the perturbation 
variables is completely determined.
From the quadratic action we derived
the conditions to avoid ghost and gradient instabilities.
They are summarized in a compact form as
\begin{equation}
{\cal F}(r)>0,\quad {\cal G}(r)>0,\quad {\cal H}(r)>0,
\end{equation}
where the three functions are dependent upon
the background metric and scalar-field profile,
as defined in Eqs.~(\ref{def-f})--(\ref{def-h}).
Given a theory and a spherically symmetric configuration,
one can judge
by checking the above three conditions
the stability of the spherically symmetric solution
under consideration and
whether or not the theory is viable.
The squared propagation speeds in the radial and angular directions are given
respectively by
$c_r^2={\cal G}/{\cal F}$ and $c_\theta^2={\cal G}/{\cal H}$,
which generically differ from the speed of light
and even do not coincide with each other.
This distinct feature in the propagation of gravitational waves is helpful to test
some of alternative theories of gravity.

In a subsequent paper~\cite{even}, we will report
the analysis of linear perturbations in the even-parity sector.
This can be done following the same line as in the present paper,
but the computation is much more involved than in the odd-parity sector
because the scalar field acquires an even-parity perturbation
as well as the metric.
It is reasonable to expect that
the quadratic action for the even-parity sector
contains two dynamical degrees of freedom,
leading to two coupled second-order master equations.

\acknowledgments
This work was supported by
JSPS Grant-in-Aid for Research Activity Start-up No.~22840011 (T.K.), 
JSPS Research Fellowships for Young Scientists (H.M.),
and JSPS Grant-in-Aid for Fellows No.~1008477 (T.S.).

\appendix
\section{Background equations}

Here we define the quantities that appear in the background Eqs.~(\ref{bgeqs}).
\begin{eqnarray}
{\cal E}_A&:=&K+B\phi'X'G_{3X}-2XG_{3\phi}
+
\frac{2}{r}\left(\frac{1-B}{r}-B'\right)  G_4
+
\frac{4B}{r}\left( 
\frac{1}{r}+\frac{X'}{X}+\frac{B'}{B} \right) X G_{4X}
+\frac{8B}{r}XX'G_{4XX}
\nonumber\\&&
-B\phi'\left(\frac{4}{r}+\frac{X'}{X}\right)
G_{4\phi}
+4X G_{4\phi\phi}
+2B\phi' \left(\frac{4}{r}-\frac{X'}{X} \right) X  G_{4\phi X}
\nonumber\\&&
+
\frac{B\phi' }{r^2}
\left[(1-3B)\frac{X'}{X}-2B' \right] X G_{5X}
-\frac{2}{r^2} B^2 \phi'XX'G_{5XX}
\nonumber\\&&
-
\frac{2}{r}\left[\frac{1+B}{r}+2B\frac{X'}{X}+B'
\right] X  G_{5\phi}
-\frac{4}{r}B\phi' XG_{5\phi\phi}
+\frac{4B}{r}\left(\frac{1}{r}-\frac{X'}{X} \right)X^2 
G_{5\phi X},
\\
{\cal E}_B&:=&K-2XK_X
+\left(\frac{4}{r}+\frac{A'}{A}\right)B \phi' XG_{3X}+2X G_{3\phi}
\nonumber\\&&
+\frac{2}{r}\left(\frac{1-B}{r}-B\frac{A'}{A}\right)G_4
-\frac{4}{r}\left(\frac{1-2B}{r}-2 B\frac{A'}{A}\right)XG_{4X}
+\frac{8B}{r}\left(\frac{1}{r}+\frac{A'}{A}\right)X^2G_{4XX}
\nonumber\\&&
-\left(\frac{4}{r}+ \frac{A'}{A}\right)B\phi'G_{4\phi}
-2\left(\frac{4}{r}+ \frac{A'}{A}\right)B\phi'XG_{4\phi X}
+\frac{B\phi'}{r^2}\left(1-5B\right)\frac{A'}{A}XG_{5X}
\nonumber\\&&
-\frac{2B^2\phi'}{r^2}\frac{A'}{A}X^2G_{5XX}
+\frac{2}{r}\left(\frac{1-3B}{r}-3 B\frac{A'}{A}\right)XG_{5\phi}
-\frac{4B}{r}\left(\frac{1}{r}+ \frac{A'}{A}\right)X^2G_{5\phi X},
\\
{\cal E}_C&:=&K+2XG_{3\phi}-X \left( B' \phi'+2B \phi'' \right) G_{3X} 
\nonumber\\&&
-\left[ \frac{1}{r} \sqrt{\frac{B}{A}} \left( r \sqrt{\frac{B}{A}} A' \right)'+\frac{B'}{r} \right]G_4
-
B\phi' \left( \frac{2}{r} + \frac{A'}{A} + \frac{B'}{B} + 2 \frac{\phi''}{\phi'} \right)
G_{4\phi}
\nonumber\\&&
+
B X \left( - \frac{A'^2}{A^2}+\frac{2}{r}\frac{B'}{B}+\frac{A'}{A}
\frac{B'}{B}+\frac{2(A'+rA'')}{rA} \right) 
\left( G_{4X}-\frac{1}{2}G_{5\phi} \right)+
BX' \left( \frac{2}{r}+\frac{A'}{A} \right) \left( G_{4X}-G_{5\phi} \right)+4X G_{4\phi\phi}
\nonumber\\&&
+ 2B\phi' \left( \frac{2}{r} + \frac{A'}{A} - \frac{X'}{X} \right) X  G_{4\phi X}
+ 2r \left( \frac{2}{r}+\frac{A'}{A} \right) XX' G_{4XX}
\nonumber\\&&
-
\frac{B^2\phi'}{2r} \left[2\frac{A''}{A}-\frac{A'^2}{A^2} + \frac{A'}{A} \left(2\frac{B'}{B}+3\frac{X'}{X} \right) \right] X G_{5X}
-B\phi' X \left( \frac{2}{r}+\frac{A'}{A} \right) G_{5\phi \phi}
\nonumber\\&&
-
\frac{B}{r} \left[ 2\frac{X'}{X}-\frac{rA'}{A} \left( \frac{2}{r}-\frac{X'}{X} \right) \right] X^2
G_{5\phi X}-\frac{B^2 \phi' A'}{rA} XX' G_{5XX},
\\
{\cal J}&:=&\phi'K_X+\left(\frac{4}{r}+\frac{A'}{A}\right)XG_{3X}-2\phi'G_{3\phi}
+2\phi'\left(\frac{1-B}{r^2}-\frac{B}{r}\frac{A'}{A}\right)G_{4X}
-\frac{ 4 B  \phi' }{r}\left(\frac{1}{r}+\frac{A'}{A}\right)XG_{4XX}
\nonumber\\&&
-2\left(\frac{4}{r}+\frac{A'}{A}\right)XG_{4\phi X}
+\frac{1-3B}{r^2}\frac{A'}{A}XG_{5X}-\frac{2B}{r^2}\frac{A'}{A}X^2G_{5XX}
\nonumber\\&&
- 2\phi' \left(\frac{1-B}{r^2}-\frac{B}{r}\frac{A'}{A}\right)G_{5\phi}
+\frac{2B\phi'}{r}\left(\frac{1}{r}+\frac{A'}{A}\right)XG_{5\phi X},
\\
{\cal S}&:=&-K_\phi+2XG_{3\phi\phi}-B\phi'X'G_{3\phi X}
-\left[\frac{B}{2}\left(\frac{A'}{A}\right)^2
+ \frac{2}{r}\left(\frac{1-B}{r}-B'\right)
-\frac{B}{2}\left\{ 2\frac{A''}{A}+\frac{A'}{A}\left(\frac{4}{r}+\frac{B'}{B}\right)
\right\} \right]G_{4\phi}
\nonumber\\&&
+B \left[
\frac{X'}{X}\left(\frac{4}{r}+\frac{A'}{A}\right)
+\frac{4}{r}\left(\frac{1}{r}+\frac{A'}{A}\right)\right] X G_{4\phi X}
+2\left(\frac{1-B}{r^2}-\frac{B}{r}\frac{A'}{A}\right)XG_{5\phi\phi}
\nonumber\\&&
- \frac{B\phi'}{r^2}\left(\frac{X'}{X}+B\frac{A'}{A}\right) X G_{5\phi X},
\end{eqnarray}



\begin{thebibliography}{99}

\bibitem{sne}
A.~G.~Riess {\it et al.},
Astron.\ J.\  {\bf 116}, 1009 (1998);
Astron.\ J.\  {\bf 117}, 707 (1999);
S.~Perlmutter {\it et al.},
Astrophys.\ J.\  {\bf 517}, 565 (1999).

\bibitem{Horndeski}
G.~W.~Horndeski, Int.\ J.\ Theor.\ Phys.\ 10 (1974) 363-384.

\bibitem{frreview}
See, {\em e.g.}, A.~De Felice and S.~Tsujikawa,
  Living Rev.\ Rel.\  {\bf 13}, 3 (2010)
  [arXiv:1002.4928 [gr-qc]].

\bibitem{Galileon}
 A.~Nicolis, R.~Rattazzi, E.~Trincherini,
  Phys.\ Rev.\  {\bf D79}, 064036 (2009).
  [arXiv:0811.2197 [hep-th]].

\bibitem{CovGali}
C.~Deffayet, G.~Esposito-Farese, A.~Vikman,
  Phys.\ Rev.\  {\bf D79}, 084003 (2009).
  [arXiv:0901.1314 [hep-th]].

\bibitem{Motohashi:2011pw} 
  H.~Motohashi and T.~Suyama,
  Phys.\ Rev.\ D {\bf 84}, 084041 (2011)
  [arXiv:1107.3705 [gr-qc]].

\bibitem{Motohashi:2011ds} 
  H.~Motohashi and T.~Suyama,
  Phys.\ Rev.\ D {\bf 85}, 044054 (2012)
  [arXiv:1110.6241 [gr-qc]].

\bibitem{GG}
  C.~Deffayet, X.~Gao, D.~A.~Steer and G.~Zahariade,
  Phys.\ Rev.\ D {\bf 84}, 064039 (2011)
  [arXiv:1103.3260 [hep-th]].

\bibitem{G2}
 T.~Kobayashi, M.~Yamaguchi, J.~Yokoyama,
  Prog.\ Theor.\ Phys.\  {\bf 126}, 511-529 (2011).
  [arXiv:1105.5723 [hep-th]].

\bibitem{gpp}
G.~Gubitosi, E.~V.~Linder,
  Phys.\ Lett.\  {\bf B703}, 113-118 (2011).
  [arXiv:1106.2815 [astro-ph.CO]].

\bibitem{gpp2}
C.~Germani, L.~Martucci, P.~Moyassari,
  Phys.\ Rev.\ D {\bf 85}, 103501 (2012)
  [arXiv:1108.1406 [hep-th]].

\bibitem{fab4}
C.~Charmousis, E.~J.~Copeland, A.~Padilla and P.~M.~Saffin,
  Phys.\ Rev.\ Lett.\  {\bf 108}, 051101 (2012)
  [arXiv:1106.2000 [hep-th]];
   C.~Charmousis, E.~J.~Copeland, A.~Padilla and P.~M.~Saffin,
  Phys.\ Rev.\ D {\bf 85}, 104040 (2012)
  [arXiv:1112.4866 [hep-th]].

\bibitem{Gao}
X.~Gao,
  JCAP {\bf 1110}, 021 (2011)
  [arXiv:1106.0292 [astro-ph.CO]];
  X.~Gao and D.~A.~Steer,
  JCAP {\bf 1112}, 019 (2011)
  [arXiv:1107.2642 [astro-ph.CO]].

\bibitem{Gao2}
  X.~Gao, T.~Kobayashi, M.~Yamaguchi and J.~Yokoyama,
  Phys.\ Rev.\ Lett.\  {\bf 107}, 211301 (2011)
  [arXiv:1108.3513 [astro-ph.CO]].

\bibitem{Tsujikawa}
A.~De Felice and S.~Tsujikawa,
  Phys.\ Rev.\ D {\bf 84}, 083504 (2011)
  [arXiv:1107.3917 [gr-qc]];
 A.~De Felice, T.~Kobayashi and S.~Tsujikawa,
  Phys.\ Lett.\ B {\bf 706}, 123 (2011)
  [arXiv:1108.4242 [gr-qc]];
 A.~De Felice and S.~Tsujikawa,
  JCAP {\bf 1202}, 007 (2012)
  [arXiv:1110.3878 [gr-qc]].

\bibitem{Rampei}
R.~Kimura, T.~Kobayashi and K.~Yamamoto,
  Phys.\ Rev.\ D {\bf 85}, 024023 (2012)
  [arXiv:1111.6749 [astro-ph.CO]];
    R.~Kimura and K.~Yamamoto,
  JCAP {\bf 1207}, 050 (2012)
  [arXiv:1112.4284 [astro-ph.CO]].

\bibitem{Regge:1957td}
  T.~Regge and J.~A.~Wheeler,
  Phys.\ Rev.\  {\bf 108}, 1063 (1957).
  
\bibitem{Zerilli:1970se}
  F.~J.~Zerilli,
  Phys.\ Rev.\ Lett.\  {\bf 24}, 737 (1970).

\bibitem{even} 
  T.~Kobayashi, H.~Motohashi, and T.~Suyama,  
  Phys.\ Rev.\ D {\bf 89}, 084042 (2014)
  [arXiv:1402.6740 [gr-qc]].

\bibitem{DeFelice:2011ka} 
  A.~De Felice, T.~Suyama and T.~Tanaka,
  Phys.\ Rev.\ D {\bf 83}, 104035 (2011)
  [arXiv:1102.1521 [gr-qc]].

\bibitem{KGB}
  C.~Deffayet, O.~Pujolas, I.~Sawicki and A.~Vikman,
  JCAP {\bf 1010}, 026 (2010)
  [arXiv:1008.0048 [hep-th]];
  O.~Pujolas, I.~Sawicki and A.~Vikman,
  JHEP {\bf 1111}, 156 (2011)
  [arXiv:1103.5360 [hep-th]];
   D.~A.~Easson, I.~Sawicki and A.~Vikman,
  JCAP {\bf 1111}, 021 (2011)
  [arXiv:1109.1047 [hep-th]].

\bibitem{GINF}
 T.~Kobayashi, M.~Yamaguchi and J.~Yokoyama,
  Phys.\ Rev.\ Lett.\  {\bf 105}, 231302 (2010)
  [arXiv:1008.0603 [hep-th]];
    K.~Kamada, T.~Kobayashi, M.~Yamaguchi and J.~Yokoyama,
  Phys.\ Rev.\ D {\bf 83}, 083515 (2011)
  [arXiv:1012.4238 [astro-ph.CO]];
    T.~Kobayashi, M.~Yamaguchi and J.~Yokoyama,
  Phys.\ Rev.\ D {\bf 83}, 103524 (2011)
  [arXiv:1103.1740 [hep-th]].

\bibitem{Yamamoto}
 R.~Kimura and K.~Yamamoto,
  JCAP {\bf 1104}, 025 (2011)
  [arXiv:1011.2006 [astro-ph.CO]];
  R.~Kimura, T.~Kobayashi and K.~Yamamoto,
  Phys.\ Rev.\ D {\bf 85}, 123503 (2012)
  [arXiv:1110.3598 [astro-ph.CO]].

\bibitem{hage}
L.~Hui and A.~Nicolis,
  Phys.\ Rev.\ Lett.\  {\bf 110}, no. 24, 241104 (2013)
  [arXiv:1202.1296 [hep-th]].

\bibitem{Pani:2009wy} 
  P.~Pani and V.~Cardoso,
  Phys.\ Rev.\ D {\bf 79}, 084031 (2009)
  [arXiv:0902.1569 [gr-qc]].

\bibitem{Germani:2011ua}
  C.~Germani and A.~Kehagias,
  Phys.\ Rev.\ Lett.\  {\bf 105}, 011302 (2010)
  [arXiv:1003.2635 [hep-ph]];
  C.~Germani and A.~Kehagias,
  JCAP {\bf 1005}, 019 (2010)
  [Erratum-ibid.\  {\bf 1006}, E01 (2010)]
  [arXiv:1003.4285 [astro-ph.CO]];
  C.~Germani and A.~Kehagias,
  Phys.\ Rev.\ Lett.\  {\bf 106}, 161302 (2011)
  [arXiv:1012.0853 [hep-ph]];
  C.~Germani and Y.~Watanabe,
  JCAP {\bf 1107}, 031 (2011)
  [Addendum-ibid.\  {\bf 1107}, A01 (2011)]
  [arXiv:1106.0502 [astro-ph.CO]].
  
\bibitem{kkred}
K.~Van Acoleyen and J.~Van Doorsselaere,
  Phys.\ Rev.\ D {\bf 83}, 084025 (2011)
  [arXiv:1102.0487 [gr-qc]].

\bibitem{bsgb}
  T.~Kobayashi and T.~Tanaka,
  Phys.\ Rev.\ D {\bf 71}, 084005 (2005)
  [gr-qc/0412139].

\end{thebibliography}
\end{document}